# Recombination of radiation defects in solid methane: neutron sources and cryo-volcanism on celestial bodies

O. Kirichek[1], E.V. Savchenko[2], C.R. Lawson[1], I.V. Khyzhniy[2], D.M. Jenkins[1], S.A. Uyutnov[2], M.A. Bludov[2], D.J. Haynes[1]

[1]ISIS Neutron and Muon Source, Science and Technology Facilities Council, Rutherford Appleton Laboratory, Didcot, OX11 0QX, UK
[2]Institute for Low Temperature Physics & Engineering NASU, Kharkiv 61103, Ukraine

oleg.kirichek@stfc.ac.uk

**Abstract**. Physicochemical properties of solid methane exposed to ionizing radiation have attracted significant interest in recent years. Here we present new trends in the study of radiation effects in solid methane. We particularly focus on relaxation phenomena in solid methane pre-irradiated by energetic neutrons and electron beam. We compare experimental results obtained in the temperature range from 10K to 100K with a model based on the assumption that radiolysis defect recombinations happen in two stages, at two different temperatures. In the case of slow heating up of the solid methane sample, irradiated at 10K, the first wave of recombination occurs around 20K with a further second wave taking place between 50 and 60K. We also discuss the role of the recombination mechanisms in "burp" phenomenon discovered by J. Carpenter in the late 1980s. An understanding of these mechanisms is vital for the designing and operation of solid methane moderators used in advanced neutron sources and could also be a possible explanation for the driving forces behind cryo-volcanism on celestial bodies.

## 1. Introduction

Solid methane is proven to be the most promising material for moderating hot neutrons produced by the target of a spallation neutron source, or in the core of nuclear reactor based neutron source. The conversion of short wavelength to long wavelength neutrons using solid methane is ~ 3.5 times more efficient than that achieved in liquid hydrogen based moderators [1]. However the history of using solid methane in cryogenic moderators knows numerous failures caused by large, seemingly spontaneous, releases of energy. This energy release is due to recombination of frozen-in products of radiolysis and the expansion of hydrogen, which builds up in the solid methane during irradiation by neutrons [2].

The physiochemical properties of solid methane and the recombination events which occur due to radiation may also prove interesting to the planetary physics community, particularly when describing comets.

According to R. Miles [3, 4] there are large cavities below a considerable fraction of the comet's surface containing methane rich solid mixtures under high pressure. This complex physicochemical system can sustain exothermic reactions at cryogenic temperatures from 40 to 200K which can provide the energy necessary for cryo-volcano eruption [3]. Energy released by the recombination of radicals generated by cosmic radiation could play significant role in this mechanism.

Here we present and discuss the results of commissioning tests of solid methane moderators at the ISIS neutron source, along with recent results obtained by experiments with solid methane samples irradiated by fast electrons. The temperature range of the moderator commissioning tests was 37 – 100K and in fast electron irradiation experiments 10 – 100K. The evidence presented here allow us to suggest that the process of radiolysis defect recombination happens in two stages, at two different temperatures. The first "low temperature" recombinations occur at around 20K, with the second "high temperature" process taking place between 50K and 60K. Consequences of this behavior for designs of solid methane moderators and cryo-volcanism on celestial bodies are also discussed.

## 2. Neutron moderators based on solid methane

The first neutron moderator based on solid methane was developed, commissioned and successfully operated for more than a decade by the IPNS neutron source at Argon National Laboratory [1, 2]. After a few initial tests it became clear that solid methane possesses superior neutronic properties to other materials, but at the same time it has reviled itself as a very challenging material from an operational point of view. As discovered by J. Carpenter in the late 1980's the accumulation of radiolysis products in low temperature (below 20K) solid methane, when exposed to a flux of neutrons, can lead to a spontaneous self-accelerated heating process, also known as the 'burp' phenomenon. It was also suggested that the 'burp' effect, together with the expansion of hydrogen (also one of the products of radiolysis), was responsible for the large number of failures of solid methane moderators. Later, solid methane exposed to neutron radiation was thoroughly studied at the IBR neutron source in Dubna [5]. The results of this research have confirmed the relevance of Carpenter's model, and give some quantitative estimation of the scale of the process.

The design of the Target Station 2 (TS2) solid methane moderator, commissioned by ISIS neutron source in 2009, was based on the conclusion that all recombination processes are completed at temperatures below 40K [1, 2, 5]. However the very first test of the moderator, operated at 38K, ended due to a burp-like effect which damaged the moderator. This observation suggested that there might be other species of radiation defects, recombining at different temperatures.

As shown in [1, 2, 5 - 7] the most substantial final products of sequential reactions of methane radiolysis are molecular hydrogen $H_2$ and ethane $C_2H_6$. Fast neutron collision with methane molecules results in a damage cascade in the irradiated methane, producing atomic hydrogen, $CH_3$ radicals and other species of radiolysis that remain dormant or 'frozen' into the cryogenic matrix of the solid methane. With rising temperature the defects start to move and (once they get close to each other) recombine, releasing significant energy in the exothermic reactions: $H + H \rightarrow H_2 + 218$ kJ/mole and $CH_3 + CH_3 \rightarrow C_2H_6 + 368$ kJ/mole [2]. These are the most probable reactions, however there are others [7] some of which lead to the production of $H_2$ and $C_2H_6$. The release of energy in the recombination events initiate heating of the cryogenic matrix and, as a result, to the acceleration of thermally activated diffusion which in turn drives the 'burp' effect. The key parameter of this process is the mobility of radiolysis products H and $CH_3$ in the cryogenic matrix of solid methane.

Another mechanism which might activate the 'burp' effect occurs when the density of frozen-in defects (at low temperatures) exceeds a critical value, when defects become close enough to each other to react. In this situation the 'burp' effect starts spontaneously, even at the lowest temperatures.

As it was suggested in [2] all recombination reactions are governed by the same activation energy for defect diffusion; $ΔE$ in equation (3) of that paper, which leads to self-accelerated heating at around 20K [2]. However during commissioning tests of the ISIS TS2 solid methane moderator we have observed a similar 'burp' effect which starts above 40K and reaches a maximum in heat release at around 60K. This allows us to suggest that the radiolysis defect recombination process happens in two stages, at different temperatures, and is therefore governed by different activation energies. In order to check this supposition we have modified the model used by Carpenter (equations (1 – 3) in [2]) based on the theory of thermal explosion [8]. We have introduced a second reaction with the rate coefficient having Arrhenius-form temperature dependence and an activation energy $\Delta E_2 \approx 200K$. Our model uses the solid methane thermal conductivity data from [9] and heat capacity from [10]. The amount of solid methane in the first prototype of the ISIS solid methane moderator was 12.5 moles. During tests the moderator was irradiated by a neutron flux ~ $10^{16}$ neutron $cm^{-2}s^{-1}$ at a temperature of 38K for periods of between 8 to 24 hours. In three instances the self-accelerating heating was triggered at ~ 50K due to warming of the moderator with the intention to anneal the solid methane. In all three cases the moderators were damaged at temperatures around 65K.

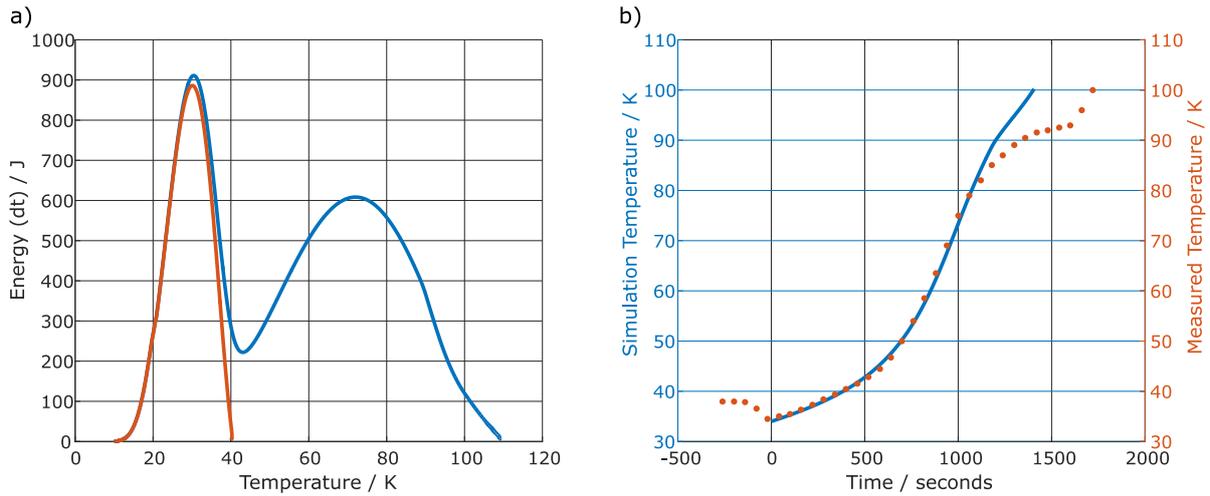

Fig 1 (a) Energy release during slow heating of the solid methane irradiated at 10K. The blue line represents our model, red line is the Carpenter model; (b) TS2 solid methane moderator test data (red dots) and time dependence of solid methane temperature obtained as an outcome of our model (blue line).

The temperatures corresponding to two maxima of energy release during slow heating of the solid methane irradiated at 10K can be determined from the Fig. 1. The blue curved line (our model) shows two distinctive maxima of energy release at 30K and 70K, contrary to the red curved line obtained from the Carpenter model showing only one maximum at 30K. One of the interesting observations is that the energy release at 30K happens much faster (hence the sharper feature) than that which happens at higher temperatures. This different dynamic could be explained by the interplay between temperature dependent parameters of the model, such as heat capacity, thermal conductivity or mobility of radiation defects.

In Fig. 1 (b) we compare experimentally obtained self-accelerating heating of the ISIS TS2 solid methane moderator (dots) with our model (solid curved line). During testing the solid methane moderator was exposed to neutron flux for 8 hours at a temperature of 38K. The proton beam was then switched of and the circulation of coolant within the moderator reduced. This causes a rise in temperature with the intention to anneal the solid methane and release hydrogen accumulated in the cryogenic matrix. This process is necessary as the buildup of defects causes a deterioration of the methane's thermal conductivity, and therefore significantly reduces the performance of the moderator over time. During this procedure initial slow heating lead to an accelerated event at around 50K and achieved the highest heating rate at around 65K. The model agrees well with experimental results and starts to deviate only above 80K where solid methane approaches its melting point at 90.7K. This deviation might be explained by not including the latent heat of solid methane melting into the model.

### 3. Relaxation phenomena in solid methane pre-irradiated with fast electrons
In the second series of experiments we study the radiation defect relaxation mechanisms in solid methane films pre-irradiated with an electron beam.

The methane solids were grown by deposition of $CH_4$ gas (of purity 99.97%) onto a cooled Cu substrate mounted in a high-vacuum chamber with a base pressure of $10^{-8}$ mbar. The technique was similar to that described in [11]. Untrapped electrons (promoted to the conduction band by heating) escape from the sample yielding thermally stimulated exo-electron emission. Stimulated currents were detected with an electrode kept at a small positive potential $V_F = +9V$ and connected to the current amplifier. The sample was irradiated with a 1keV electron beam of 1.2 mAcm$^{-2}$ current density in dc regime. The sample heating under electron beam did not exceed 0.5K. The cathodoluminescence (CL) spectra were detected concurrently in the visible range and in the vacuum ultraviolet (VUV) using two spectrometers. The CL of solid methane steeply increased upon switching on the beam, reached a maximum and then decayed exponentially, indicating radiation-induced chemical modification of the film. Upon completion of the electron beam exposure an after-emission of electrons was detected, pointing to an electrostatic charging of the $CH_4$ film. When the after-emission current had decayed to essentially zero, measurements of thermally stimulated relaxation emissions were performed in the temperature range 5 – 80K. During a controlled warm-up of the solid methane sample (with a heating rate of 5K min$^{-1}$) three thermally stimulated effects were monitored: VUV and visible photon emission, an exo-electron emission and the ice's particle ejection. In view of the absence of thermally stimulated luminescence (TSL) from pure solid $CH_4$ only the emissions of electrons and particles have been measured in a correlated fashion.

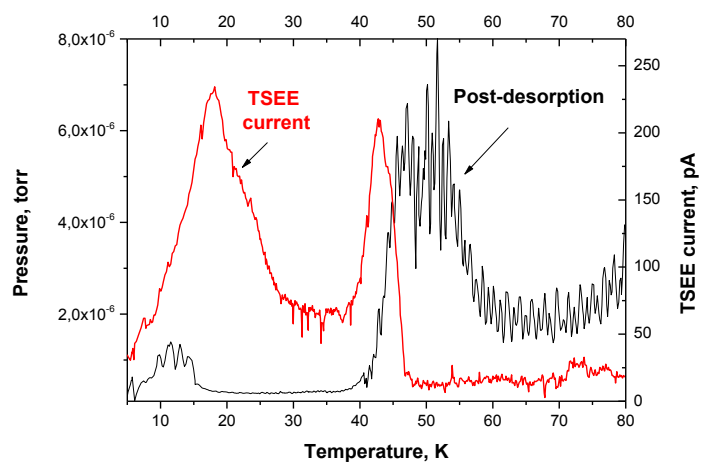

Fig 2 Temperature dependences of thermally stimulated exo-electron emission TSEE (red line) and post-desorption of solid methane or products of radiolysis (black line).

In Fig.2 we present temperature dependences of thermally stimulated exo-electron emission (TSEE) (red line) and post-desorption of solid methane or products of radiolysis (black line). The TSEE signal (two peaks at about 18 and 43K) was detected from solid methane for the first time. The first peak related to defects of structure represents in fact a multi-peak curve with unresolved structure. A rough estimation of the averaged activation energy $E_{act}$ by the half-width method [12] gives $E_{act}$~ 10meV. $E_{act}$ for the second peak at 43K is about 70 meV. A check on the presence of TSL made in the visible and VUV range verified its absence in pure solid methane. First peak of particle ejection (post-desorption) was found in solid methane at temperatures much lower than the characteristic sublimation temperatures. Another double peak in post-desorption signal was observed at higher temperatures – between 45 and 60K.

## 4. Discussion and conclusions
The radiation defect recombination process observed around 20K is well known and has been a subject of thorough investigation since the 1980s [1, 2, 5 - 7]. However the second exothermal recombination around 50K has manifested itself quite unexpectedly and has given rise to a number of questions about the physical mechanisms behind this process. It is worth mentioning that anomalies in the same temperature range have also been observed in other parameters of solid methane. As shown in [13, 14] the transport and mechanical properties of solid methane undergo significant changes at temperatures around 50K [14]: in the 50K – 90K temperature range solid methane is soft and sticky, but below 50K solid methane loses its stickiness and ductility and behaves more or less like an ordinary window glass. In [13] it was recently suggested that this change might be caused by quantum effects of rotational degrees of freedom and their collectivization. However, irrespective to the microscopic mechanism which governs this process, it is clear that dynamics of crystal defects such as dislocations, vacancies or impurities experience dramatic change in this temperature range.

According to the scenario suggested here, in the low temperature limit only 'light' defects (such as atomic hydrogen, protons or electrons) are capable of thermal diffusion, and therefore recombination. All massive defects, such as $CH_3$, remain motionless. If the temperature starts to increase above 20K the recombination rate of 'light' mobile defects rises, following an Arrhenius thermal activation dependence. In the case where the accumulated number of defects is significant, this process could self-accelerate: Triggering the 'burp' phenomenon. After all mobile defects have recombined the heating driven by 'light' defect recombination stops. If the temperature increases further and reaches 50K, where the 'heavy' defects begin to move, the self-accelerated heating process repeats itself again, with a different thermal activation energy in the Arrhenius temperature dependence. We accept that the suggested scenario is quite speculative and more information, particularly related to microscopic dynamics of radiation defects is required. Nevertheless, the suggested model describes test results reasonably well and allows us to propose some modifications to the design of solid methane moderators.

The first, and most important, issue is that the higher temperature self-accelerated heating happens in the temperature range 40K to 70K where solid methane undergoes dramatic thermal expansion [15]. The expansion (if it happens quickly) might generate significant pressure on the walls of the moderator vessel, which could lead to structural damage. In order to prevent this buildup of pressure it may be possible to rapidly warm the coolant circulated through the cooling loop, prior to beginning of the annealing process. This quick heating could allow warming up of the solid methane layer close to the surface of the heat exchanger. The rest of the bulk solid methane is expected to remain at low temperature due to the poor thermal conductivity of solid methane [9]. At temperatures above 50K solid methane becomes soft [14] and can be squeezed out of the warm layer by the rising pressure, releasing the pressure on the moderator walls. The second option would be to never fill the moderator with methane up to 100% and always leave some empty space, where soft methane from the warm layer can be squeezed out. The last approach has been already incorporated into the ISIS TS2 solid methane moderator operational procedure.

In addition to this, in the case of the ISIS TS2 moderator, it was decided to increase the thickness of the aluminum walls of the vessel from 3 mm to 10 mm. Such radical design change has completely eliminated the possibility of moderator damage, but at the same time significantly reduced the cold neutron flux.

In addition to the application of defect recombination models in solid methane, Carpenter has previously suggested that the 'burp' phenomenon might play some role in generating jets, such as those observed from comets [2]. Comets spend a significant amount of time in the low temperature environment of the Kuiper Belt and the Oort Cloud, exposed to the continuous radiation of outer space. This allows bulk solid methane inclusions (whose existence is suggested in [3]) to accumulate a significant amount of radiation defects, which in turn could lead to spontaneous self-accelerated heating. However the role of this phenomenon in cryo-volcanism on comets was downplayed later [3] mainly on the basis of differences in temperature ranges: the 'burp' effect takes place around 20K, but the temperature of a comet is never thought to drop below 30K, with cryo-eruptions usually happening above 50K. According to R. Miles [3, 4] this cryo-volcanism is mostly driven by the release of the enthalpy of solution, also known as 'heat of solution' [3]. However, to start the process the system would still require an initial heat release in combination with the generation of pressure, in many cases around 1 kPa. The higher temperature self-accelerated heating effect observed here at 50K provides the necessary heating and pressure to trigger the eruption of a cryo-volcano. Furthermore, we would suggest that radiation defect recombination in water ice [5] might also play noticeable role in cryo-volcanism on outer celestial bodies.

We conclude that the process of radiolysis defect recombination happens in two stages at different temperatures. First the "lower temperature" recombination (also known as the 'burp' phenomenon) occur at around 20K and the second "higher temperature" self-accelerated heating process takes place between 50 and 60K. We also discussed the consequences of this observation in the design and operation of solid methane moderators, as well as its role as a possible trigger of cryo-volcanism on celestial bodies such as comets.

**Acknowledgments**
We thank our colleagues J. M. Carpenter, E. P. Shabalin and S. A. Kulikov for their active interest and useful discussions and M. I. Bagatskii for sharing the solid methane thermodynamic data with us. The work done at ISIS was supported by the project TUMOCS. This project has received funding from the European Union's Horizon 2020 research and innovation program under the Marie Skłodowska-Curie Grant No. 645660.